\documentclass[twocolumn,amsmath,amssymb,nofootinbib,prd]{revtex4}

\def\gsim{ \lower .75ex \hbox{$\sim$} \llap{\raise .27ex
\hbox{$>$}} }
\def\lsim{ \lower .75ex \hbox{$\sim$} \llap{\raise .27ex
\hbox{$<$}} }
\usepackage{graphicx}
\usepackage{amssymb}
\usepackage{bbm}

\def\ba{\begin{eqnarray}}
\def\ea{\end{eqnarray}}
\def\be{\begin{equation}}
\def\ee{\end{equation}}
\def\ben{\begin{equation} \nonumber}
\def\een{\end{equation}}
\def\baray{\begin{eqnarray*}}
\def\earay{\end{eqnarray*}}
\def\M{{\cal M}}
\def\delr{\delta_{D-r}({\cal M}^r)}

\def\Z{{\mathbbm Z}}
\def\RP{{\mathbbm R}{\mathrm{P}^3}}
\def\R{{\mathbbm R}}

\def\d{{\rm d}}
\def\ti{{t_{\rm i}}}
\def\tf{{t_{\rm f}}}
\def\ff{{f_\phi}}
\def\nh{{\hspace{-1mm}}}

\begin{document}

\title{Chaotic brane inflation}
\author{Benjamin Shlaer}
\affiliation{Institute of Cosmology, Department of Physics and Astronomy\\Tufts University, Medford, MA  02155, USA}

\begin{abstract}\noindent
We illustrate a framework for constructing models of chaotic inflation where the inflaton is the position
of a D3-brane along the universal cover of a string compactification.  
In our scenario, a brane rolls many times around a nontrivial one-cycle, thereby unwinding a Ramond-Ramond flux.  These ``flux monodromies" are similar in spirit to the monodromies of Silverstein, Westphal, and McAllister,
and their four-dimensional description is that of Kaloper and Sorbo.  Assuming moduli stabilization is rigid enough, the large-field inflationary potential is protected from radiative corrections by a discrete shift symmetry.
\end{abstract}


\maketitle
\section{Introduction}
Perhaps the simplest phenomenological model of inflation \cite{Guth:1980zm,Linde:1981mu,Albrecht:1982wi} is due to Linde's monomial potential \cite{Linde:1983gd}, which undergoes what is called ``chaotic inflation" due to its expected behavior on large scales.  Chaotic inflationary models are not obviously natural in the context of effective field theory precisely because of the requirement that the potential be sufficiently flat over super-Planckian field distances.  In the quadratic model, the
inflaton mass must be of order $10^{-5}M_{\rm P}$, and higher-order terms in the potential must remain subdominant for field values as large as $10 M_{\rm P}$, which requires a functional fine-tuning from an effective field theory point of view.  

An elegant solution to this problem
was presented in Refs.~\cite{Kaloper:2008fb,Kaloper:2011jz}, whereby an axion ``eats" a three-form potential, and so acquires a mass \cite{Dvali:2001sm}.  The axion potential is purely quadratic, being protected from radiative corrections by the underlying shift symmetry.  Aside from the simplicity of the model, chaotic inflation is interesting because of its distinct
phenomenological predictions; it is capable of sourcing significant primordial tensor perturbations, which may be detectible in the cosmic microwave background.

Here we will find a stringy realization of large-field inflation.  As in brane inflation \cite{Dvali:1998pa,Jones:2002cv,Burgess:2001fx}, the inflaton represents the position of a D3-brane in a six-dimensional compactification manifold, assumed to be sufficiently stable.  The potential felt by the D3-brane due to the five-form field strength $F_5$ gives rise to the four-dimensional effective potential of the inflaton.  Crucially, this field strength depends not just on the location of the D3-brane(s), but also on their history, i.e.,  the number of times they have traversed any nontrivial one-cycles of the compactification manifold.  This ``flux wrapping" will allow for the possibility of large-field brane inflation, which cannot otherwise exist \cite{Chen:2006hs,Baumann:2006cd} because increasing the field range typically requires increasing the compactification volume, which in turn increases the four dimensional Planck mass.  

It should be pointed out that there are already a few stringy realizations of large-field inflation, e.g. Refs.~\cite{Silverstein:2008sg, McAllister:2008hb,Dong:2010in}, as well as Refs.~\cite{Brandenberger:2008kn,Avgoustidis:2006zp,Avgoustidis:2008zu}.  After this paper was completed, we learned of related work on unwinding fluxes \cite{Kleban:2011cs}, and their application to inflation  \cite{D'Amico:2012sz,D'Amico:2012ji}.

In the probe approximation, the potential felt by a D3 must be
exactly periodic, just as for an axion.  Furthermore, the five-form flux takes quantized values over the five-cycle which is dual to the one-cycle.  Assuming rigid moduli stabilization, the flux potential is exactly quadratic in the {\em discrete} flux winding $\oint_{\M^5}F_5$.  By turning on the coupling of the D3 to the background five-form flux, the periodicity is lifted, and the discrete flux becomes a {\em continuous} parameter, contributing an exactly quadratic term to the potential.  We now illustrate this with a simple example.

\section{Charges in compact spaces with nontrivial first homology}
As a warm-up example, let us consider a single electron and positron in the compact
space $S^2\times S^1$.  The action and equations of motion are given by
\ba
S &=& \int_{\mathbbm{R}\times S^2\times S^1}\hspace{-11mm}\tfrac{1}{2} \d A_1\wedge\ast \d A_1 + eA_1\wedge\delta_3(\M^1),\\
\d\ast \d A_1 &=& e\,\delta_3(\M^1),
\ea
where $\M^1$ is the oriented world lines of the charges.
Integration and differentiation of the equations of motion require that the point-particle current $\delta_3(\M^1)$ satisfy 
\be
\int_{S^2\times S^1} \delta_3(\M^1) = 0\quad\mbox{and}\quad \d\delta_3(\M^1) = 0,
\ee
or equivalently (see Appendix), 
\be\label{eq:ppconstraints}
 S^2\hspace{-1mm}\times \hspace{-1mm}S^1\cap \,\M^1=\, 0 \quad \mbox{and} \quad \partial \M^1 = \emptyset,
\ee
respectively.  We abuse the notation $\cap$ to mean both the intersection and the winding number of the intersection, so the left-hand side of Eq.~(\ref{eq:ppconstraints}) should be read as stating that there are equally many positive points as negative points in the total intersection.  

These equations simply state that no net charge can occupy a compact space, and electric current is conserved.
We can either ensure that $\M^1$ has no {\em net} time-like winding (as we have done), or add a diffuse  background ``jellium" charge to the action.  A homogenous jellium contribution is just proportional to the spatial volume form,
\be
\delta_3(\M^1) \,\,\to\,\, \delta_3(\M^1) - n \frac{{\rm vol}_3(S^2\times S^1)}{\int {\rm vol}_3(S^2\times S^1)},
\ee
with $n = \left( S^2\hspace{-1mm}\times \hspace{-1mm}S^1\cap \,\M^1\right)$.   The uniform charge density cancels the tadpole.

Let us imagine that $\M^1$ represents a single positive charge and a single negative charge.  We can compute the potential between them by finding the Green's function on this space.  We expect the usual Coulomb interaction to be modified by two effects: 
\begin{itemize}
\item Because the space is compact the potential will not be Coulombic at large distances.
\item Because the space has nontrivial first homology $H^1(S^2\times S^1) = \mathbbm{Z}$, the field strength will not be single-valued, but will depend on the winding of the particles' paths. 
\end{itemize}
It is the latter effect which we find useful here, as it enables one to change the electric flux on
the $S^1$.  It is straightforward to calculate the difference in flux caused by transporting {\em one} of the two charges around the $S^1$.  The transport of one of the particles around the one-cycle means that $\M^1$ acquires winding number equal to one.  The flux on the $S^1$ is measured by choosing a fixed time and $z$ coordinate, and then integrating the dual field strength $\tilde{F_2}=\ast\,\d A_1$ over the $S^2$.  

To calculate the change in flux caused by a single winding of a particle,
let us define a 3-manifold (with boundary) $\M^3$ which spans an interval in time $[\ti, \tf]$ times the full $S^2$ cycle.  Then
\baray
\int_{\M^3}\d\ast \d A_1 &=& \int_{\partial \M^3}\nh\nh \ast \d A_1 = \left.\int_{S^2}\nh\ast\d A_1\right|_{\ti}^{\tf}\\
=e\int_{\M^3}\delta_3(\M^1) &=& e\,\,\M^1 \cap \M^3 = e,
\earay
and so
\be
\Delta F_2 = \frac{e\ast {\rm vol}_2(S^2)}{\int {\rm vol}_2(S^2)}.
\ee
Thus the electric field in the $z$ direction changes by one unit each time a particle is transported around the circle in the $z$ direction.  A simple interpretation of this is that the charge drags the field lines around the cycle.  

We can immediately write down the homological piece of the potential.  If the metric is given by $\d s^2 = -\d t^2 + R^2 \left(\d\theta^2 + \sin^2\theta \d\phi^2\right) + \d z^2$, with $z \equiv z + L$, then
\be
A_1(\Delta z) = \frac{\Delta z}{L}\frac{ez}{4\pi R^2} \d t \quad+\quad \mbox{single-valued part}
\ee
where $\Delta z$ is the $z$ separation of the two charges as measured on the universal cover.
The flux part of the electron potential is thus 
\be\label{eq:potential}
V(z) = \frac{e^2z^2}{4\pi R^2L} = \frac{e^2z^2}{V_\perp},
\ee
where $V_\perp = \int {\rm vol}_3(S^2\times S^1)=4\pi R^2L$.

The flux potential cancels the jellium \cite{Shandera:2003gx,Rabadan:2002wy} contribution.  We can think of the jellium term in the potential
as arising due to the finite compactification volume.  
A jellium term is required in the potential felt by a probe charge, since the field strength is then single-valued.  But transport of a {\em physical} charge around a nontrivial cycle does not leave the field strength invariant, and so the probe charge is an inadequate description.  

In a sense, one can say that the configuration space of charges and flux is not simply the product of the compact manifold and its first homology, but rather is a nontrivial fibration:  one can change the flux by transporting charges around the one-cycles associated with them.  

Although the potential is exactly quadratic classically (and even in perturbation theory), there are nonperturbative corrections.  
Pair production will eventually discharge any potential exceeding twice the electron mass. This is such a slow enough process that we can safely ignore it.  Furthermore, adiabatic motion of a charge will never be able to wind more than one unit of flux because of avoided level crossing \cite{Kaloper:2011jz}.  This is not a problem except on timescales long compared to $m^{-1} \exp(m L)$, where $m$ is the electron mass.

\section{Ingredients for Chaotic Brane Inflation}
The ingredients we will need is an F-theory compactification of Type IIB string theory which
contains at least one mobile D3-brane.  Furthermore, the six-dimensional transverse space must have a nontrivial first homology, i.e., $H^1(\M^6) = \Z$ or $\Z_N$.  Because the D3 moduli in the direction of the nontrivial one-cycles are lifted at tree level, these models may lack supersymmetry.  All closed string moduli must be sufficiently stabilized, in order that inflation may take place in this background.  It is further necessary that the periodic portion of the potential be flat enough that the full potential has only a single minimum.  In the probe approximation, the D3 has a discreet ``shift symmetry" associated with transport about the one-cycle, but this may not be sufficient to guarantee local flatness.  As illustrated before, the inflationary potential exists due to nontrivial winding of the five-form flux about the homological one-cycle.  The D3-brane moves classically through this cycle to unwind the flux.  We will assume that the moduli stabilization is rigid enough to ignore the backreaction of the dynamical flux.  This assumption is generically false in known warped flux compactifications \cite{Kachru:2003sx,McAllister:2005mq,Baumann:2006th}, but such effects may actually flatten the potential \cite{Dong:2010in}.

The potential induced by brane monodromy is
\be
V(z) = \frac{\mu_{{\rm D}3}^2z^2}{M_{10,{\rm P}}^8L^6}
\ee
where $L^6$ is the volume of the compact space, $M_{10,{\rm P}}$ is the ten-dimensional reduced Planck mass, and $\mu_{\rm D3} = M_{10,{\rm P}}^4$ is the D3 charge.  We assume the string coupling to be of order unity.
In terms of the four-dimensional reduced Planck mass, $M^2_{\rm P} = M_{10,{\rm P}}^8L^6$, and for a canonically normalized inflaton $\phi = \sqrt{\mu_{{\rm D}3}} z$, we find the potential
\be
V(\phi) = \frac{\phi^2}{\mu_{{\rm D}3}L^6} = \frac{\phi^2}{M_{\rm P}L^3}.
\ee

To achieve reasonable density perturbations, the quadratic model needs the inflaton mass to
obey
\be
m_\phi^2 \approx 10^{-11}M_{\rm P}^2,
\ee
which requires the compactification scale to be
\be
L \approx \frac{10}{M_{10,{\rm P}}}\approx \frac{10^4}{M_{\rm P}}.
\ee  
In terms of the inflaton, this scale corresponds to a field distance
\be
2\pi \ff = \sqrt{\mu_{\rm D 3}}L = \sqrt{\frac{M_{\rm P}}{L}}.
\ee
Hence, successful large-field inflation will require the brane to undergo of order a few thousand revolutions, so any model must have first Homology
large enough to permit this, i.e. 
\be
H_1(\M^6) = \Z \quad {\rm or} \quad \Z_N
\ee
with $N \gtrsim 2\times10^3$.  This rather large number can be relaxed by no more than two orders of magnitude by allowing the size of the one-cycle to be much larger than the natural scale $\sqrt[6]{L^6}$.

In the ${\mathbbm Z}$ case,
the quadratic approximation for $V(\phi)$ must eventually break down, due to backreaction.  If we assume the modulus $L$ is very heavy, it can be written as $L(\phi)$, which grows as more flux is wound on the transverse space.  If this is the only effect of backreaction, the inflaton potential is flattened at large flux values, and reaches a maximum if $\d L(\phi)/\d\phi$ exceeds  $\tfrac{2}{3} L(\phi)/\phi$.  

However, a steepening \cite{Dong:2010in} of the inflaton potential could instead occur due to kinetic coupling between the inflaton and $L(\phi)$, say of the form
\be
\left(\frac{L(\phi)}{L(0)}\right)^n \frac{\dot\phi^2}{2},
\ee
for sufficiently negative $n$.
We will assume that the kinetic coupling is subdominant, and hence that backreaction leads to a flattening of the inflaton potential at sufficiently large field values $\phi \sim N \ff$.  
Qualitatively speaking, a potential which is quadratic at small field values, and flat at large field values can be thought of as approximately sinusoidal over the range $\left|\phi\right| \lesssim N\ff$.

The ${\mathbbm Z}_N$ case has extended periodicity $z \equiv z + N L$, and so the homological part of the potential in {\em each} of the above cases is approximately given by
\be\label{eq:natural}
V(\phi) \approx \Lambda^4\left[1-\cos\left(\frac{\phi}{N\ff}\right)\right],
\ee
with 
\be
\Lambda = \frac{\sqrt{N}}{\sqrt[4]{2\pi^2} L}, \quad \ff = \frac{\sqrt{M_{\rm P}}}{2\pi\sqrt{L}},
\ee
where $N$ represents either the backreaction scale or the size of the homology group.
This scenario could be called
natural brane inflation, following Refs.~\cite{Freese:1990rb,Freese:2004un}, although it avoids the problems associated with large axion decay constants $\ff$ by the appearance of the large factor $N$ in the potential of Eq.~(\ref{eq:natural}), allowing $\ff\ll M_{\rm P}\ll N \ff$.  Alternative approaches to this problem can be found in Refs.~\cite{Dimopoulos:2005ac,ArkaniHamed:2003wu}.

We additionally require the single-valued part of the potential
\be
V_{\rm s.v.}(\phi) \approx \lambda^4 \cos\left(\phi/\ff\right)
\ee
 to be relatively flat, meaning $\lambda \lesssim 1/L$.

To achieve 60 $e$-folds of slow-roll inflation, we must arrange the scalar field $\phi$ to initially have a super-Planckian vacuum expectation value, $\phi \gtrsim 15 M_{\rm P}$.  The Hubble scale during inflation is then $H \approx 10^{-5}M_{\rm P}$, which is almost two orders of magnitude below the Kaluza-Klein scale $2\pi/L$.  However, the four-dimensional potential will be of order $V \approx 10^{-10}M_{\rm P}^4$, which exceeds the tension of a D3-brane by two orders of magnitude, opening the possibility for brane tunneling \cite{Brown:2007zzh} or nucleation \cite{Brown:1988kg}.  Because these are slow processes,  our description remains valid.  Indeed, brane nucleation\footnote{This possibility is thoroughly considered in Refs.~\cite{D'Amico:2012sz,D'Amico:2012ji}.} could give rise to the mobile inflaton, although inflation will then end with brane-antibrane annihilation, but unlike Ref.~ \cite{Brown:2008ea}, the bubble need not self-annihilate until after many laps are completed.   The final D3-$\overline{{\rm D}3}$ annihilation will result in the formation of a cosmic-string network \cite{Sarangi:2002yt}.

\section{Discussion}
We have provided a simple framework for large-field brane inflation.  To construct realistic models, a number of hurdles must first be addressed, the most significant of which is moduli stabilization.  However, because our framework relies on a nontrivial first homology group, much of the progress made on moduli stabilization of warped compactifications does not apply here.
  Another potential difficulty may arise in obtaining a flat enough periodic portion of the brane potential.  If the modulation of the quadratic piece is too large, there may not be a long enough slow-roll trajectory.  On the other hand, a periodic modulation of the inflaton potential can lead to detectable non-Gaussanity in the cosmic microwave background \cite{Chen:2008wn}.  Finally, it is unlikely that supersymmetry can be unbroken in the models considered here, since the D3 moduli receive an explicit mass, rather than from a spontaneous uplifting, say by the introduction of antibranes.

Nevertheless, a number of intriguing features arise here, foremost being a UV description of large-field inflation.  The monodromies of this framework are extremely easy to visualize, being simply the motion of a (point-like) brane around a one-cycle.  By traversing the cycle (perhaps several thousand
times) a Ramond-Ramond flux is unwound, realizing either chaotic or natural inflation, both of which predict significant tensor modes in the cosmic microwave background.

\begin{acknowledgments}
We thank Daniel Baumann, Xingang Chen, Liam McAllister, Enrico Pajer, Lorenzo Sorbo, Alexander Westphal, and Xi Dong for helpful conversations.  Funding was provided through NSF grant PHY-1213888.
\end{acknowledgments}

\begin{appendix}

\section{The de Rham delta function}
Here we review a simple notation \cite{Shlaer:2006ti} appropriate for calculating the effects of localized sources coupled to gauge potentials.  The new object is a singular differential form which we call the ``de Rham delta function."
\subsection{Definition}
On a $D$-dimensional oriented manifold $\M^D$ with $\M^r \subseteq \M^D$ an oriented submanifold of dimension $r$, we define the de Rham delta function $\delr$ as follows:
\be
\int_{\M^D} \hspace{-4mm} C_r\wedge\delr \quad = \quad \int_{\M^D\cap \M^r}\hspace{-10mm}C_r\quad\quad ,
\label{definition}
\ee
where the pullback is implicit on the rhs.
The subscripts denote the order for differential forms, and superscripts denote the dimension for manifolds.
Stokes' theorem then implies 
\baray
\int_{\partial \M^D}\hspace{-5mm}C_{r-1}\wedge\delr &=&  \int_{\M^D}\hspace{-3mm}\left[ \d C_{r-1}\wedge\delr\right. \\ &&\left.\, +\,  (-1)^{r-1} C_{r-1}\wedge \d\delr\right] \\
&=& \int_{\partial(\M^D\cap\M^r)}\hspace{-14mm}C_{r-1}\qquad +\\&&(-1)^{r-1}\hspace{-1mm}\int_{\M^D}\hspace{-4mm}C_{r-1}\wedge \d\delr,
\earay
and so
\be
\d \delr = (-1)^r \delta_{D - r + 1}(\partial\M^r),
\label{stokes}
\ee
where we have used the fact that  $\partial(\M^r\cap\M^s) = (\partial \M^r\cap\M^s) \,\cup \,(-1)^{D-r}
(\M^r\cap\partial\M^s)$.  Here $\cup$ is essentially the group sum of $r$-chains in $\M^D$.  This definition of $\cup$ is equivalent
to 
\be
\delta_{D-r}(\M^r \cup  \M^{\prime r}) = \delr + \delta_{D-r}(\M^{\prime r}).
\label{sum}
\ee

Following the definition we also find
\baray
 \int_{\M^D\cap\M^s\cap\M^r}\hspace{-16mm}C_{r+s-D} &=& \int_{\M^D\cap\M^s}\hspace{-10mm}C_{r+s-D}\hspace{-1mm}\wedge\hspace{-1mm}\delr,\\
&=& \int_{\M^D}\hspace{-4mm}C_{r+s - D}\hspace{-.5mm}\wedge\hspace{-2mm}\ \delr\hspace{-1mm}\wedge\hspace{-1mm}\delta_{D - s}(\M^s),
\earay
which leads to the relation
\be
\delr\wedge\delta_{D-s}(\M^s) = \delta_{2D - r - s}(\M^s\cap\M^r).
\label{wedge}
\ee

This identity illuminates some generic features of submanifolds.
\begin{itemize}
\item The intersection of an $r$- and an $s$-dimensional submanifold in $D$ dimensions
will generally be of dimension $r+s-D$.
\item When the previous statement does not hold, integration on the intersection must vanish.  This
is because the intersection is not stable under infinitesimal perturbation (and not transversal).

\item When two submanifolds each have odd codimension, the orientation of their intersection
flips when the order of the manifolds is reversed.  This is consistent with the Leibniz rule for the
boundary operator given below Eq.~(\ref{stokes}).  As an example of this, consider two 2-planes in three dimensions, whose
intersection is a line.  The orientation of each plane is characterized by a normal vector, and
the antisymmetric cross product of these is used to determine the orientation of the line of intersection.

\item We should think of $\cap$ as being the oriented intersection operation from intersection homology which makes
the above properties automatic.  It is stable under infinitesimal perturbation of either submanifold.

\end{itemize}

\subsection{Coordinate representation}
The coordinate representation of $\delr$ is straightforward in coordinates where
the submanifold is defined by the $D-r$ constraint equations,
\be
q \in \M^r  \Rightarrow \lambda^i(x^1[q], ... ,x^D[q]) = 0,
\label{lambda}
\ee
with $ i = 1 ... D-r$, via
\be
\delr = \delta^{(D-r)}(\lambda^i) \, \, \d\lambda^1\hspace{-1.0mm}\wedge....\wedge\hspace{-.2mm}\d\lambda^{D-r},
\label{coordinate}
\ee
where $\delta^{(D-r)}$ is the usual $(D-r)$-dimensional Dirac delta function.
The well-known transformation properties of the Dirac delta function make this object
automatically a differential form.  (Thus the only meaningful zeros of the $\lambda^i$ are transversal zeros, i.e., those where $\lambda^i$ changes sign in any neighborhood of the zero.)  If a submanifold is $D$-dimensional, then the corresponding
de Rham delta function is simply the characteristic function, $\delta_0(\M^{\prime D}) = \chi_{{\M^{\prime D}}}$, with
\be
\chi_{\M^{\prime D}}(q) = \left\{
\begin{array}{ll}
1& \quad q \in \M^{\prime D},\\
0& \quad q \notin \M^{\prime D}.
\end{array}
\right.
\ee
One may describe submanifolds with boundary by multiplication with this scalar de Rham delta function using Eq.~(\ref{wedge}).
As an example, if $\M^1$ is the positive $x$ axis in ${\mathbbm R}^3$, then
\be
\delta_2(\M^1) = \delta(y)\delta(z)\Theta(x)\d y\wedge \d z,
\label{xaxis}
\ee
where the characteristic function is the Heaviside function $\Theta(x)$. 
Notice that the orientation of this submanifold has been chosen to be along the $+x$ direction, 
consistent with Eq.~(\ref{stokes}) and the fact that its boundary is {\em minus} the point at the origin.

The $\lambda^i$ do not need to be well defined on the entire manifold, and in fact they only need to be defined at all in a neighborhood of $\M^r$.  Thus despite its appearance in Eq.~(\ref{coordinate}), 
$\delr$ is not necessarily a total derivative.   
If all of the $\lambda^i$ are well defined everywhere, then $\M^r$ is an algebraic variety.  By
Eq.~(\ref{stokes}) and Eq.~(\ref{coordinate}) it can be seen that all algebraic varieties can be
written globally as boundaries.  It may be that the $\lambda^i$ are well defined only in a 
neighborhood of $\M^r$, in which case $\M^r$ is a submanifold.  Near points not on $\partial \M^r$
we may think of $\M^r$ locally as a boundary, just as we may think of a closed differential form as locally
exact.  Nonorientable submanifolds will correspond to constraints that may be
double valued, that is $\lambda^i$ may return to minus itself upon translation around the submanifold.
We considered such cases in Ref.~\cite{Shlaer:2006ti}.  

Another important case occurs when $\M^r$ is only an
immersion, i.e., it intersects itself.  The $\lambda^i$ are path dependent here, as well.  Consider
the immersion $S^1 \subset \R^2$ defined by the constraint $\lambda = 2 \arcsin(y) - \arcsin(x) = 0$.
This looks like a figure-eight centered on the origin of the plane.
Clearly $\lambda$ is multivalued, and to get a complete figure-eight requires summing over two
branches of $\lambda$.  We suppress the sum in Eq.~(\ref{coordinate}).  Notice that the figure-eight
immersion satisfies
\baray
\delta_1(S^1)\wedge\delta_1(S^1) = \delta_2(0) - \delta_2(0) = 0.
\earay
The self-intersection of this immersion is twice the point at the origin, but since the orientation (sign)
of the intersection is negative for exactly one of the two points of intersection, the total self-intersection
vanishes.  For even-codimemsion immersions, the sum over branches allows for nonzero self-intersection from cross terms.  By the antisymmetry of the wedge product, one may show that the self-intersection of an immersion of odd codimension will always vanish, assuming $\M^D$ is orientable.   
One may also compute the $n$th self-intersection of an immersion via
\baray
\delta_{n(D-r)}(\cap^n \M^r) = \bigwedge_n\delr.
\earay

\subsection{Geometry}
If we introduce a metric on our manifold, we can measure the volume of our submanifolds
with the volume element from the pull-back metric.  We use $\ast_r$ to denote the Hodge star on $\M^r$, so the induced volume element is $\ast_r1$.  Looking at the
coordinate definition of $\delr$ given in Eq.~(\ref{lambda}), we see that each of the $\lambda^i$ is
constant along the submanifold, and so $\d\lambda^i$ is orthogonal to the submanifold.  Thus, a
good candidate
for a volume element on $\M^r$ is $\ast (\d\lambda^1\wedge...\wedge \d\lambda^{D-r})$.
To remove the rescaling redundancy in the $\lambda^i$'s we may divide by $||\bigwedge_i \d\lambda^i||$, where the norm of a differential form $C_r$ is defined by the scalar
\baray
||C_r|| = \sqrt{\left| \ast \left(C_r\wedge\ast C_r \right) \right|}.
\earay
Hence, we write
\be
\ast_r1 = \frac{\ast \delr}{||\delr||},
\label{volume}
\ee
where the pull-back is implicit on the rhs.  Despite its appearance,  $\frac{\ast \delr}{||\delr||}$ is an $r$-form living in all $D$ dimensions, although it is only well defined near $\M^r$.  This is because in Eq.~(\ref{volume}), the Dirac delta
function coefficients cancel, leaving dependence
only on the smooth $\lambda^i$s.  Because of this, we may define $\d\frac{\ast\delr}{||\delr||}$ using
the full $D$-dimensional exterior derivative.  This is well defined on  $\M^r$, and when restricting to points on the submanifold,
\be
\d \ast\frac{ \delr}{||\delr||} = 0\quad \Longleftrightarrow \quad \M^r \quad{\textrm{is extremal}}.
\ee

From Eq.~(\ref{volume}) it is evident that 
\baray
||\delr|| = \ast_r \ast \delr.
\earay
  More generally, the action by this Hodge star can be
rewritten as
\be
\ast_r F_s = \ast \frac{ F_s\wedge \delr}{||\delr||},
\label{hodge}
\ee
where $F_s$ is an $s$-form living on $\M^r$ and the pull-back is implicit on the rhs.

\subsection{Topology}
One feature of the de Rham delta function is that Poincar\'e duality becomes manifest.
Here we assume $\M^D$ is orientable and compact with no boundary.  Then Eq.~(\ref{stokes}) tells us that
$\delr$ is closed if and only if $\M^r$ is a cycle, and $\delr$ is exact if $\M^r$ is
a boundary.  To complete this correspondence between the $r$th homology and the
$(D-r)$th de Rham cohomology of $\M^D$, we need to show that
\be
\delr = \d f_{D-r-1} \quad \Longrightarrow \quad \M^r = \partial \M^{r+1}.
\ee

But this is not true when torsion is present.  Consider the manifold $\RP =  {\rm SO(3)}$ which has
a single nontrivial one-cycle $\zeta^1$, i.e. $H_1(\RP;\Z) = \Z_2$.   Since the group sum of two
of these cycles is trivial, they must form a boundary, $\zeta^1 + \zeta^1 = \partial \M^2$
and so $2\delta_2(\zeta^1) = \d\delta_1(\M^2)$ which means
\be
\delta_2(\zeta^1) = \frac{1}{2}\d\delta_1(\M^2).
\ee
Since $\zeta^1$ is not a boundary, it is false to claim that all exact de Rham delta functions are Poincar\'e dual to boundaries.
Only by using real coefficients can we make the statement
 that  
 \be 
 \M^r \simeq \M^{\prime r} \quad \Longleftrightarrow \quad [\delr] \simeq [\delta_{D-r}(\M^{\prime r})].
 \ee
De Rham's theorem gives us the isomorphism between homology and cohomology
\baray
H_r(\M^D;{\mathbbm R}) \cong H^r(\M^D;{\mathbbm R}),
\earay
and Poincar\'e duality asserts that
\baray
H_r(\M^D;{\mathbbm R}) \cong H_{D-r}(\M^D;{\mathbbm R}).
\earay
The de Rham delta function provides the isomorphism
\be
H_r(\M^D;{\mathbbm R}) \cong H^{D-r}(\M^D;{\mathbbm R}).
\ee
In fact, if the cohomology basis is chosen such that
\baray
\int_{\M_{(j)}^{D-r}}\omega^{(i)}_{D-r} = \delta^{ij},
\earay
then
\be
[\delta_{D-r}(\widetilde{\M}^r_{(i)})] = [\omega^{(i)}_{D-r}],
\ee
where $\widetilde{\M}_{(i)}^r$ is the Poincar\'e dual of $\M^{D-r}_{(i)}$, and together they satisfy
\baray
\widetilde{\M}_{(i)}^r\cap\M_{(j)}^{D-r} = \delta_{ij},
\earay
i.e., their net intersection is a single positive point if $i = j$, and is empty otherwise.

\end{appendix}


\begin{thebibliography}{99}

\bibitem{Guth:1980zm}
  A.~H.~Guth,
  ``The Inflationary Universe: A Possible Solution To The Horizon And Flatness Problems,''
  Phys.\ Rev.\  D {\bf 23}, 347 (1981).

\bibitem{Linde:1981mu}
  A.~D.~Linde,
  ``A New Inflationary Universe Scenario: A Possible Solution Of The Horizon, Flatness, Homogeneity, Isotropy And Primordial Monopole Problems,''
  Phys.\ Lett.\  B {\bf 108}, 389 (1982).
  
\bibitem{Albrecht:1982wi}
  A.~J.~Albrecht and P.~J.~Steinhardt,
  ``Cosmology For Grand Unified Theories With Radiatively Induced Symmetry Breaking,''
  Phys.\ Rev.\ Lett.\  {\bf 48}, 1220 (1982).

\bibitem{Linde:1983gd} 
  A.~D.~Linde,
  ``Chaotic Inflation,''
  Phys.\ Lett.\ B {\bf 129}, 177 (1983).


\bibitem{Kaloper:2008fb} 
  N.~Kaloper and L.~Sorbo,
  ``A Natural Framework for Chaotic Inflation,''
  Phys.\ Rev.\ Lett.\  {\bf 102}, 121301 (2009)
  [arXiv:0811.1989 [hep-th]].
  
\bibitem{Kaloper:2011jz} 
  N.~Kaloper, A.~Lawrence and L.~Sorbo,
  ``An Ignoble Approach to Large Field Inflation,''
  JCAP {\bf 1103}, 023 (2011)
  [arXiv:1101.0026 [hep-th]].
  
\bibitem{Dvali:2001sm} 
  G.~R.~Dvali and A.~Vilenkin,
  ``Field theory models for variable cosmological constant,''
  Phys.\ Rev.\ D {\bf 64}, 063509 (2001)
  [hep-th/0102142].


\bibitem{Dvali:1998pa}
  G.~R.~Dvali and S.~H.~H.~Tye,
  ``Brane inflation,''
  Phys.\ Lett.\  B {\bf 450}, 72 (1999)
  [arXiv:hep-ph/9812483].

\bibitem{Jones:2002cv} 
  N.~T.~Jones, H.~Stoica and S.~H.~H.~Tye,
  ``Brane interaction as the origin of inflation,''
  JHEP {\bf 0207}, 051 (2002)
  [hep-th/0203163].

\bibitem{Burgess:2001fx}
  C.~P.~Burgess, M.~Majumdar, D.~Nolte, F.~Quevedo, G.~Rajesh and R.~J.~Zhang,
  ``The Inflationary Brane-Antibrane Universe,''
  JHEP {\bf 0107}, 047 (2001)
  [arXiv:hep-th/0105204].


\bibitem{Chen:2006hs} 
  X.~Chen, S.~Sarangi, S.~-H.~Henry Tye and J.~Xu,
  ``Is brane inflation eternal?,''
  JCAP {\bf 0611}, 015 (2006)
  [hep-th/0608082].
  
\bibitem{Baumann:2006cd}
  D.~Baumann and L.~McAllister,
  ``A Microscopic Limit on Gravitational Waves from D-brane Inflation,''
  Phys.\ Rev.\  D {\bf 75}, 123508 (2007)
  [arXiv:hep-th/0610285].

\bibitem{Silverstein:2008sg}
  E.~Silverstein and A.~Westphal,
  ``Monodromy in the CMB: Gravity Waves and String Inflation,''
  Phys.\ Rev.\  D {\bf 78}, 106003 (2008)
  [arXiv:0803.3085 [hep-th]].

\bibitem{McAllister:2008hb}
  L.~McAllister, E.~Silverstein and A.~Westphal,
  ``Gravity Waves and Linear Inflation from Axion Monodromy,''
  arXiv:0808.0706 [hep-th].
  

\bibitem{Dong:2010in}
  X.~Dong, B.~Horn, E.~Silverstein, A.~Westphal,
  ``Simple exercises to flatten your potential,''
  [arXiv:1011.4521 [hep-th]].

  

\bibitem{Brandenberger:2008kn}
  R.~H.~Brandenberger, A.~Knauf and L.~C.~Lorenz,
  ``Reheating in a Brane Monodromy Inflation Model,''
  JHEP {\bf 0810}, 110 (2008)
  [arXiv:0808.3936 [hep-th]].

\bibitem{Avgoustidis:2006zp}
  A.~Avgoustidis, D.~Cremades and F.~Quevedo,
  ``Wilson line inflation,''
  Gen.\ Rel.\ Grav.\  {\bf 39}, 1203 (2007)
  [arXiv:hep-th/0606031].

\bibitem{Avgoustidis:2008zu}
  A.~Avgoustidis and I.~Zavala,
  ``Warped Wilson Line DBI Inflation,''
  JCAP {\bf 0901}, 045 (2009)
  [arXiv:0810.5001 [hep-th]].

\bibitem{Kleban:2011cs} 
  M.~Kleban, K.~Krishnaiyengar and M.~Porrati,
  ``Flux Discharge Cascades in Various Dimensions,''
  JHEP {\bf 1111}, 096 (2011)
  [arXiv:1108.6102 [hep-th]].

\bibitem{D'Amico:2012sz} 
  G.~D'Amico, R.~Gobbetti, M.~Schillo and M.~Kleban,
  ``Inflation from Flux Cascades,''
  arXiv:1211.3416 [hep-th].

\bibitem{D'Amico:2012ji} 
  G.~D'Amico, R.~Gobbetti, M.~Kleban and M.~Schillo,
  ``Unwinding Inflation,''
  arXiv:1211.4589 [hep-th].



\bibitem{Rabadan:2002wy}
  R.~Rabadan and F.~Zamora,
  ``Dilaton tadpoles and D-brane interactions in compact spaces,''
  JHEP {\bf 0212}, 052 (2002)
  [arXiv:hep-th/0207178].


\bibitem{Shandera:2003gx}
  S.~Shandera, B.~Shlaer, H.~Stoica and S.~H.~H.~Tye,
  ``Inter-brane interactions in compact spaces and brane inflation,''
  JCAP {\bf 0402}, 013 (2004)
  [arXiv:hep-th/0311207].






\bibitem{Kachru:2003sx}
  S.~Kachru, R.~Kallosh, A.~Linde, J.~M.~Maldacena, L.~P.~McAllister and S.~P.~Trivedi,
  ``Towards inflation in string theory,''
  JCAP {\bf 0310}, 013 (2003)
  [arXiv:hep-th/0308055].

\bibitem{McAllister:2005mq}
  L.~McAllister,
  ``An inflaton mass problem in string inflation from threshold corrections  to volume stabilization,''
  JCAP {\bf 0602}, 010 (2006)
  [arXiv:hep-th/0502001].

\bibitem{Baumann:2006th}
  D.~Baumann, A.~Dymarsky, I.~R.~Klebanov, J.~M.~Maldacena, L.~P.~McAllister and A.~Murugan,
  ``On D3-brane potentials in compactifications with fluxes and wrapped D-branes,''
  JHEP {\bf 0611}, 031 (2006)
  [arXiv:hep-th/0607050].





\bibitem{Freese:1990rb} 
  K.~Freese, J.~A.~Frieman and A.~V.~Olinto,
  ``Natural inflation with pseudo - Nambu-Goldstone bosons,''
  Phys.\ Rev.\ Lett.\  {\bf 65}, 3233 (1990).


\bibitem{Freese:2004un} 
  K.~Freese and W.~H.~Kinney,
  ``On natural inflation,''
  Phys.\ Rev.\ D {\bf 70}, 083512 (2004)
  [hep-ph/0404012].

\bibitem{Dimopoulos:2005ac} 
  S.~Dimopoulos, S.~Kachru, J.~McGreevy and J.~G.~Wacker,
  ``N-flation,''
  JCAP {\bf 0808}, 003 (2008)
  [hep-th/0507205].


\bibitem{ArkaniHamed:2003wu} 
  N.~Arkani-Hamed, H.~-C.~Cheng, P.~Creminelli and L.~Randall,
  ``Extranatural inflation,''
  Phys.\ Rev.\ Lett.\  {\bf 90}, 221302 (2003)
  [hep-th/0301218].


\bibitem{Brown:2007zzh} 
  A.~R.~Brown, S.~Sarangi, B.~Shlaer and A.~Weltman,
  ``A Wrinkle in Coleman-De Luccia,''
  Phys.\ Rev.\ Lett.\  {\bf 99}, 161601 (2007)
  [arXiv:0706.0485 [hep-th]].

\bibitem{Brown:1988kg} 
  J.~D.~Brown and C.~Teitelboim,
  ``Neutralization of the Cosmological Constant by Membrane Creation,''
  Nucl.\ Phys.\ B {\bf 297}, 787 (1988).


\bibitem{Brown:2008ea}
  A.~R.~Brown,
  ``Boom and Bust Inflation: a Graceful Exit via Compact Extra Dimensions,''
  Phys.\ Rev.\ Lett.\  {\bf 101}, 221302 (2008)
  [arXiv:0807.0457 [hep-th]].


\bibitem{Sarangi:2002yt} 
  S.~Sarangi and S.~H.~H.~Tye,
  ``Cosmic string production towards the end of brane inflation,''
  Phys.\ Lett.\ B {\bf 536}, 185 (2002)
  [hep-th/0204074].

\bibitem{Chen:2008wn} 
  X.~Chen, R.~Easther and E.~A.~Lim,
  ``Generation and Characterization of Large Non-Gaussianities in Single Field Inflation,''
  JCAP {\bf 0804}, 010 (2008)
  [arXiv:0801.3295 [astro-ph]].


\bibitem{Shlaer:2006ti}
  B.~S.~P.~Shlaer, (Ph.D. dissertation, Cornell University, 2006)
  ``Cosmic Strings In The Brane World,''



\end{thebibliography}
\end{document}